\documentclass[aps,prl,twocolumn,superscriptaddress,showpacs,amsmath,amssymb]{revtex4-1}

\usepackage{graphicx}

\begin{document}

\title{Competition between superconductivity and magnetic/nematic order
as a source of anisotropic superconducting gap in underdoped Ba$_{1-x}$K$_{x}$Fe$_{2}$As$_{2}$}

\author{H. Kim}
\affiliation{Ames Laboratory, Ames, Iowa 50011, USA}

\author{M. A. Tanatar}
\affiliation{Department of Physics \& Astronomy, Iowa State University, IA 50011,
USA}

\affiliation{Ames Laboratory, Ames, Iowa 50011, USA}

\author{W. E. Straszheim}
\affiliation{Ames Laboratory, Ames, Iowa 50011, USA}

\author{K.~Cho}
\affiliation{Ames Laboratory, Ames, Iowa 50011, USA}

\author{J.~Murphy}
\affiliation{Department of Physics \& Astronomy, Iowa State University, IA 50011,
USA}

\author{N.~Spyrison}
\affiliation{Department of Physics \& Astronomy, Iowa State University, IA 50011,
USA}

\author{J.-Ph.~Reid}
\affiliation{Departement de physique \& RQMP, Universite de Sherbrooke, Sherbrooke,
Canada}

\author{ Bing Shen}
\affiliation{Center for Superconducting Physics and Materials,
National Laboratory of Solid State Microstructures $\&$ Department of Physics, Nanjing University, Nanjing 210093, China}

\author{Hai-Hu Wen}
\affiliation{Center for Superconducting Physics and Materials,
National Laboratory of Solid State Microstructures $\&$ Department of Physics, Nanjing University, Nanjing 210093, China}

\author{R. M. Fernandes}

\affiliation{School of Physics and Astronomy, University of Minnesota, Minneapolis,
MN 55455, USA}

\author{R. Prozorov}

\email[Corresponding author: ]{prozorov@ameslab.gov}

\affiliation{Department of Physics \& Astronomy, Iowa State University, IA 50011,
USA}

\affiliation{Ames Laboratory, Ames, Iowa 50011, USA}

\date{9 June 2014}

\begin{abstract}
The in-plane London penetration depth, $\Delta\lambda(T)$, was measured
using a tunnel diode resonator technique in single crystals of Ba$_{1-x}$K$_{x}$Fe$_{2}$As$_{2}$
with doping levels $x$ ranging from heavily underdoped, $x$=0.16
($T_{c}$=7~K) to nearly optimally doped, $x$= 0.34 ($T_{c}=$39
K). Exponential saturation of $\Delta\lambda(T)$ in the $T\to0$
limit is found in optimally doped samples, with the superfluid density
$\rho_{s}(T)\equiv(\lambda(0)/\lambda(T))^{2}$ quantitatively described
by a self-consistent $\gamma$-model with two nodeless isotropic superconducting
gaps. As the doping level is decreased towards the extreme end of
the superconducting dome at $x$=0.16, the low-temperature behavior
of $\Delta\lambda(T)$ becomes non-exponential and best described
by the power-law $\Delta\lambda(T)\propto T^{2}$, characteristic
of strongly anisotropic gaps. The change between the two regimes happens
within the range of coexisting magnetic/nematic order and superconductivity,
$x<0.25$, and is accompanied by a rapid rise in the absolute value
of $\Delta\lambda(T)$ with underdoping. This effect, characteristic
of the competition between superconductivity and other ordered states,
is very similar to but of significantly smaller magnitude than what
is observed in the electron-doped Ba(Fe$_{1-x}$Co$_{x}$)$_{2}$As$_{2}$
compounds. Our study suggests that the competition between superconductivity
and magnetic/nematic order in hole-doped compounds is weaker than
in electron-doped compounds, and that the anisotropy of the superconducting
state in the underdoped iron pnictides is a consequence of the anisotropic
changes in the pairing interaction and in the gap function promoted
by both magnetic and nematic long-range order.
\end{abstract}


\maketitle

\section{introduction}

The experimental determination of the symmetry of the superconducting
gap is important to unravel the mechanism of superconductivity in
iron-based superconductors \cite{Mazin2010Nature,Wang2011,reviews_pairing}.
Measurements of the London penetration depth \cite{Gordon2009,Gordon2009a,Martin2010},
thermal conductivity \cite{Tanatar2010,Reid2010} and specific heat
\cite{Gofryk2011,Bud'ko2009,Hardy2010} in electron doped Ba(Fe$_{1-x}$Co$_{x}$)$_{2}$As$_{2}$
(BaCo122) suggest that the superconducting gap changes significantly
with doping, developing nodes at both overdoped and underdoped dome
edges \cite{Reid2010,Hirschfeld2010physics,Maiti2011}. This doping
evolution is very similar to what is observed in another electron-doped
family, NaFe$_{1-x}$Co$_{x}$As \cite{KyuilNaFeAs,M2s} and LiFeAs
\cite{Borisenko2010,Inosov2010,Kim2011,Tanatar2011}. It is also consistent
with the predicted dependence of the gap function with doping in the
$s^{+-}$ model \cite{Chubukov2009,Maiti2011}. On the other hand,
nodal behavior is observed at all doping levels in isovalent- substituted
BaFe$_{2}$(As$_{1-x}$P$_{x}$)$_{2}$ (BaP122) \cite{Hashimoto2010a}.
This remarkable contrast in two systems that share the same parent
compound prompts a detailed study of the hole doped Ba$_{1-x}$K$_{x}$Fe$_{2}$As$_{2}$
(BaK122) materials.

In BaK122, a full isotropic gap has been reported in compositions
close to optimal \cite{Ding2008,Nakayama2011,Li2011,Martin2009,Hashimoto2009,Luo2009,Dong2010},
whereas strongly overdoped compositions with $x\approx1$ display
nodal superconductivity \cite{Fukazawa2009,Dong2010,Hashimoto2010,Hardy2013,Matsudaoverdoped}.
Although these observations suggest a similar trend as compared to
the electron-doped BaCo122 materials \cite{ReidSUST}, there has been
no systematic studies of the superconducting gap structure in the
underdoped BaK122 so far. In this doping regime, superconductivity
coexists and competes with long-range magnetic/nematic order \cite{Chubukov2009,Fernandes2010},
making this an ideal system to investigate the rich interplay between
these ordered states \cite{Chubukovanomaly,Fernandes2014}. Interestingly,
a sizeable asymmetry between the normal state properties of the electron-
and hole-doped materials is observed in such underdoped regime \cite{Blomberg2013,Elena,BaNa}.

\begin{figure*}[ht]
\includegraphics[width=0.9\linewidth]{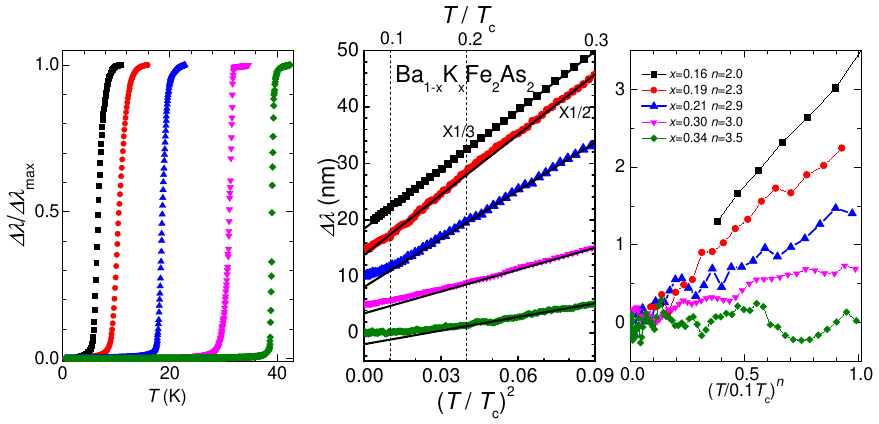} \caption{ (a) Normalized $\Delta\lambda(T)$ for samples of Ba$_{1-x}$K$_{x}$Fe$_{2}$As$_{2}$
with $x$=0.16, 0.19, 0.21, 0.3 and 0.34 (left to right). Middle panel
(b) shows zoom of the actual $\Delta\lambda(T)$ plotted vs $(T/T_{c})^{2}$
for a range $T/T_{c}\leq$0.3. Right panel (c) shows the same data
over the lowest temperature range $T/T_{c}\leq$ 0.1, plotted as a
function of $(T/0.1T_{c})^{n}$ with $n$ 2.0, 2.3, 2.9, 3.0 and 3.5
$\pm$0.1 for $x$= 0.16, 0.19, 0.21, 0.3 and 0.34, respectively.
Note the systematic decrease of $\Delta\lambda(T)$ on approaching
optimal doping.}

\label{fig1}
\end{figure*}

In this work we study the evolution of the temperature dependence
of the in-plane London penetration depth, $\Delta\lambda(T)$, in
high quality single crystals of Ba$_{1-x}$K$_{x}$Fe$_{2}$As$_{2}$
across the underdoped region of the phase diagram $0.16\leq x\leq0.34$.
We find that the optimally doped samples show a weak exponential temperature
dependence in the $T\to0$ limit, suggesting nodeless isotropic gaps.
This conclusion is consistent with the temperature dependence of the
superfluid density, which can be well fitted using the self-consistent
$\gamma$-model with two full gaps in the clean limit \cite{Kogan2009}.
In contrast, the lowest-$T_{c}$ samples, deep in the underdoped regime,
show a strong power-law temperature dependence of $\Delta\lambda(T)$
in the low-$T$ limit, typical of strongly anisotropic gaps. The onset
of this behavior coincides with the onset of the coexistence between
the superconducting and magnetic/nematic phases, indicating that the
anisotropic changes in the pairing interaction arising from this coexistence
play a major role in determining the gap structure in the underdoped
regime. The magnitude of $\Delta\lambda(0.25T_{c})$, which serves
as a proxy of the magnitude of the zero-temperature penetration depth,
shows a rapid rise in the range of coexisting magnetic/nematic order
and superconductivity. Comparison with the penetration depth data
in the electron-doped counterpart Ba(Fe$_{1-x}$Co$_{x}$)$_{2}$As$_{2}$
reveals that the rapid rise of $\Delta\lambda$ in the coexistence
state is similar in both materials, as expected for competing electronic
ordered states \cite{Fernandes_Schmalian2,Vavilov_penetration_depth,Levchenko13,Sachdev13}.
However, the increase in $\Delta\lambda$ is almost three times larger
in electron-doped materials, suggesting that the competition between
magnetic/nematic order and superconductivity is weaker in the hole-doped
materials. Interestingly, this electron-hole asymmetry inside the
superconducting state correlates with the asymmetric behavior of the
normal state properties, in particular the nematic susceptibility,
as measured by the in-plane resistivity anisotropy \cite{Blomberg2013}
and by elastic constant measurements \cite{Boehmer2013}, and pseudogap features in the inter-plane resistivity \cite{Tanatar2010b,Tanatar2011BaKAniso}.

\section{experimental}

The growth and characterization of single crystals of BaK122 used
in this study is described in detail in previous reports \cite{Tanatar2011BaKAniso,Luo2008}.
In brief, measurements were performed on pre-screened crystals with
sharp superconducting transitions and individually measured chemical
compositions with wavelength dispersive x-ray spectroscopy (WDS) in
\textit{JEOL JXA-8200} electron microprobe. The composition was measured
for 12 points per single crystal and averaged, yielding statistical
errors of compositional measurement of $\pm$0.005. The London penetration
depth $\Delta\lambda(T)$ was measured using the tunnel-diode resonator
technique \cite{Degrift1975,Prozorov2006,Prozorov2000} in our He$^{3}$
and dilution refrigerator set-ups; details of the calibration procedure
and data analysis can be found in Ref.~\onlinecite{Prozorov2011}.

\section{Results and discussion}

In Fig.~\ref{fig1}(a) we show the variation of the London penetration
depth $\Delta\lambda(T)$ from the base temperature to $T_{c}$ for
five compositions of Ba$_{1-x}$K$_{x}$Fe$_{2}$As$_{2}$ spanning
from $x$=0.16 ($T_{c}=7$ K, edge of the superconducting dome) to
$x$=0.34 ($T_{c}=39$ K, nearly optimally doped). The data are normalized
by $\Delta\lambda(T_{c})$ and reveal the high quality of our single
crystals as evidenced by the sharpness of the superconducting transitions
and the absence of any additional features. In Fig.~\ref{fig1}(b)
we present the same data plotting the actual $\Delta\lambda(T)$ as
function of the reduced temperature $T/T_{c}$ for $0<T/T_{c}\leq0.3$.
This is the characteristic temperature range in which the superconducting
gap of single-band superconductors can be considered constant, and
in which the temperature dependence of $\Delta\lambda(T)$ reflects
the nodal structure of the gap. In the clean limit, $\Delta\lambda(T)$
is expected to depend exponentially on temperature in nearly-isotropic
nodeless superconductors, whereas a linear-in-$T$ behavior is expected
for superconductors with line nodes. Because the experimental verification
of the exponential dependence is difficult due to noise in the data,
the standard procedure is to fit the data to a power-law function,
$\Delta\lambda(T)=AT^{n}$. In this case, exponents $n\gtrsim3$ usually
indicate nodeless gap, whereas $n\lesssim2$ indicate strong gap anisotropies
-- either due to nodes in the presence of impurity scattering or due
to very deep gap minima \cite{KyuilNaFeAs}. In Fig.~\ref{fig1}(b)
the data are plotted as function of $(T/T_{c})^{2}$. Several features
can be noticed: first, the samples at the very edge of the dome, $x$=0.16,
display a temperature dependence very close to $T^{2}$, signaling
a sizeable gap anisotropy. As optimal doping is approached, the magnitude
of $\Delta\lambda(0.3T_{c})$ dramatically decreases and the $\Delta\lambda(T)$
curves progressively flatten at low temperatures. This flattening
signals a full-gap state, though we note that data for $T/T_{c}>0.15$
approximately follow a $T^{2}$ behavior.

\begin{figure}[tb]
\includegraphics[width=1\linewidth]{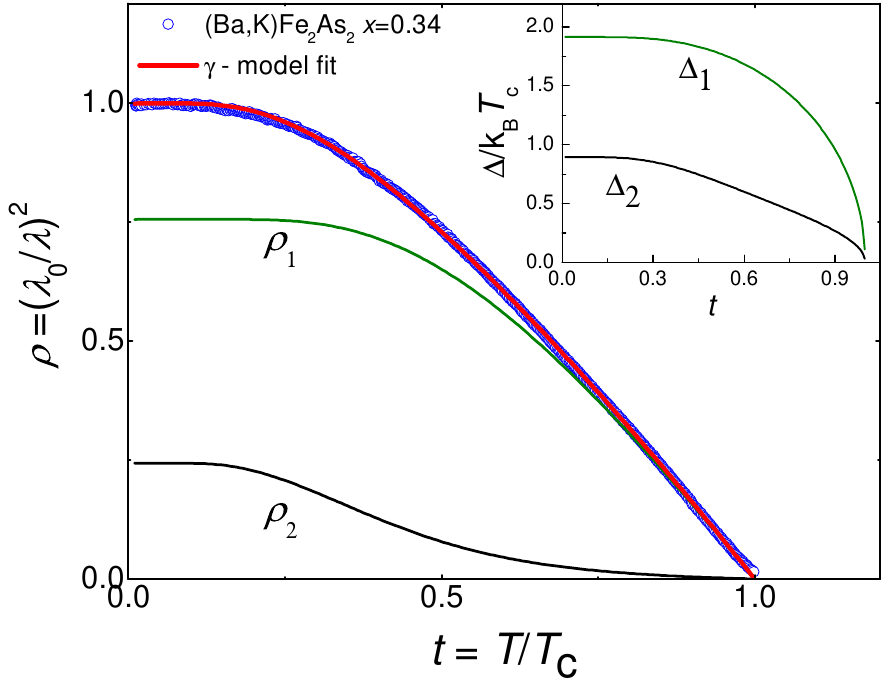} \caption{ (Color online) Superfluid density, $\rho_{s}(T)\equiv(\lambda(0)/\lambda(T))^{2}$,
of the optimally doped sample Ba$_{1-x}$K$_{x}$Fe$_{2}$As$_{2}$,
$x$=0.34, calculated from the data of Fig.~\ref{fig1} using $\lambda(0)$=200~nm
\cite{Li2008} (open symbols). The solid red line on top of the data
is the fit using the self-consistent two-gap $\gamma$-model, with
the black and green curves in the main panel denoting the partial
superfluid densities $\rho_{1}$ (larger gap) and $\rho_{2}$ (smaller
gap). The temperature dependence of the corresponding gaps $\Delta_{1}$
and $\Delta_{2}$ is shown in the inset.}

\label{fig2}
\end{figure}

To shed light on the behavior at optimal doping, in Fig.~\ref{fig2}
we show the superfluid density, $\rho_{s}(T)=\lambda^{2}(0)/\lambda^{2}(T)$
of the $x$=0.34 sample. Here, $\lambda(T)=\Delta\lambda(T)+\lambda(0)$
is obtained by using $\lambda(0)=200$ nm \cite{Li2008} as the value
of the London penetration depth in the $T\to0$ limit. The superfluid
density $\rho_{s}(T)$ shows a clear saturation at low temperatures,
evidencing a full-gap superconducting state, similar to the data of
Fig.~\ref{fig1}(b). A more detailed analysis of the $\rho_{s}(T)$
data was made using the clean-limit $\gamma-$model to fit the data
\cite{Kogan2009}, as shown by the red solid line in the same figure.
Here $\rho_{s}=\gamma\rho_{1}+(1-\gamma)\rho_{2}$; the partial superfluid
densities $\rho_{1}$ and $\rho_{2}$ are shown in the main panel
of Fig.~\ref{fig2}, whereas the superconducting gaps $\Delta_{1}(T)$
and $\Delta_{2}(T)$ are shown in the inset of Fig.~\ref{fig2}.
The estimated gap values in the $T\to$0 limit are 6.5 meV and 3.3
meV, the larger gap being in reasonable agreement with the value of
$\sim6$ meV found from specific heat measurements \cite{Mu2009Cp}.
This analysis implies that the small superconducting gap, $\Delta_{2}$,
is strongly temperature dependent even for $T<0.3T_{c}$, and the
characteristic behavior can be found only at temperatures at least
two times lower than $0.3T_{c}$. In Fig.~\ref{fig1}(c) we show
the data over the temperature range $0<T/T_{c}\leq0.1$, presented
as a power-law function $T^{n}$ of the reduced temperature, with
exponent $n$ as shown in the main panel.

\begin{figure}[tb]
\includegraphics[width=0.8\linewidth]{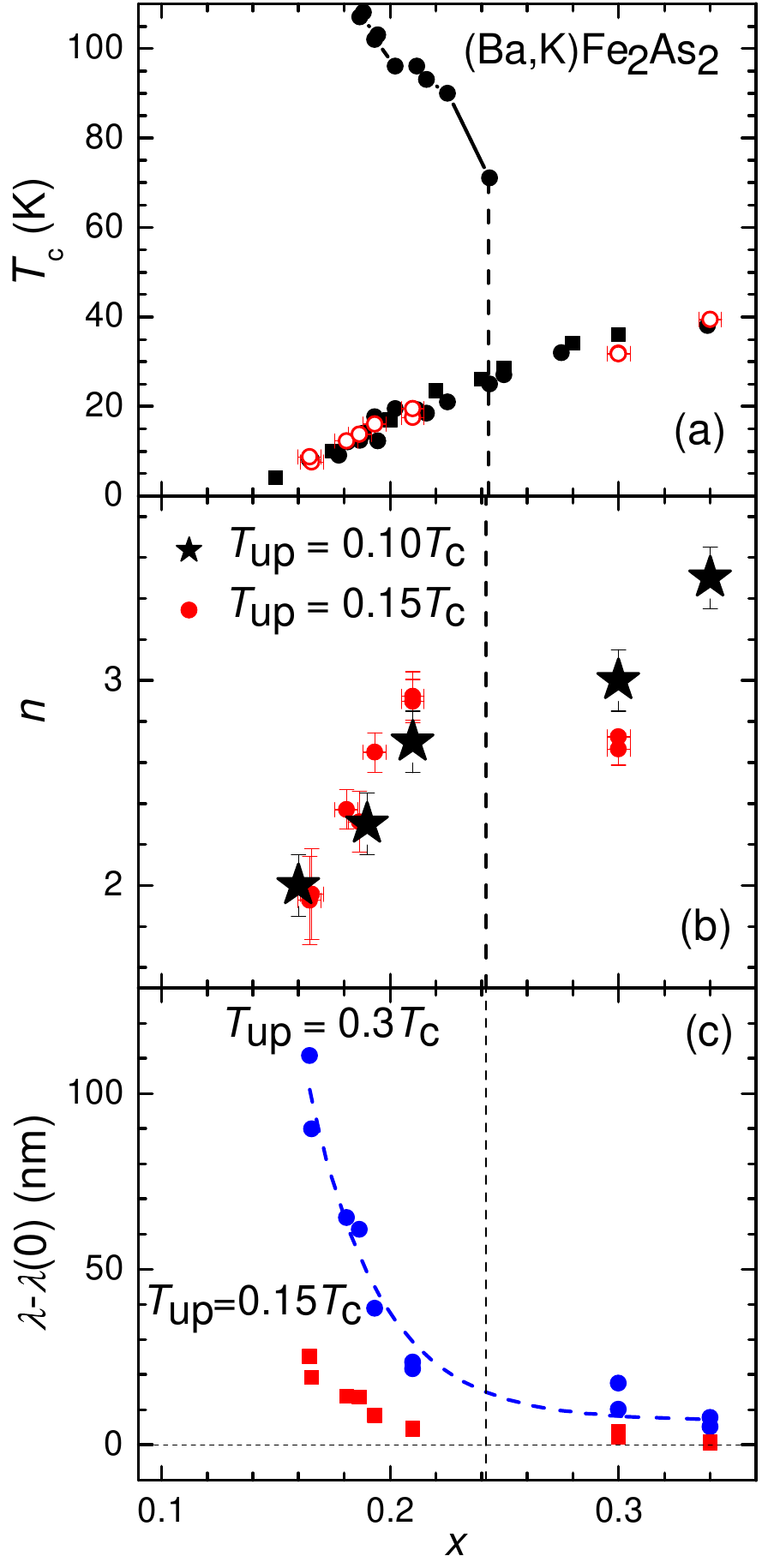} \caption{(a) Doping phase diagram of Ba$_{1-x}$K$_{x}$Fe$_{2}$As$_{2}$
as determined from resistivity and TDR measurements on single crystals
\cite{Tanatar2011BaKAniso} (black squares and dots), matching well
the results from neutron scattering and magnetization measurements
on polycrystalline samples \cite{Rotter2008,Avci2011}. Dashed line
shows an extrapolation of the orthorhombic/magnetic transition lines,
$T_{sm}(x)$ to $T\to0$. The middle panel (b) shows the doping evolution
of the exponent of the power-law function $n$ as determined from
the data analysis in the temperature range $T/T_{c}\leq$0.1 (black
stars) and $T/T_{c}\leq$0.15 (red circles). The bottom panel shows
the doping evolution of the magnitude of the variation of the London
penetration depth at low temperatures, $\Delta\lambda(T_{up})$ with
$T_{up}$=0.15$T_{c}$ (red squares) and $T_{up}$=0.3$T_{c}$ (blue
circles).}

\label{fig3}
\end{figure}

In Fig.~\ref{fig3} we summarize the doping evolution of the London
penetration depth as found in our study. For reference in the top
panel Fig.~\ref{fig3}(a) we show the doping phase diagram as determined
from our TDR and resistivity measurements \cite{Tanatar2011BaKAniso},
which are in good agreement with the phase diagram determined from
neutron scattering and magnetization data on polycrystalline samples
of Avci \textit{et al.} \cite{Avci2011}. This analysis reveals two
clear trends: (i) The exponent $n$ of the power-law temperature dependence
of $\Delta\lambda$, Fig.~\ref{fig3}(b), as determined from the
data analysis for $0<T\leq T_{up}$ with $T_{up}$=0.1$T_{c}$ and
$T_{up}$=0.15$T_{c}$, decreases from $n$=3.5 for $x$=0.34 (which
is technically indistinguishable from an exponential dependence) to
$n=2$ for $x$=0.16. (ii) The actual variation of the London penetration
depth $\Delta\lambda(T_{up})$ with $T_{up}$=0.3$T_{c}$ and $T_{up}$=0.15$T_{c}$
-- which mimics the doping dependence of the zero-temperature penetration
depth -- strongly increases in the same doping regime.

Both effects are more prominent for doping levels $x\leq0.21$, where
magnetism and nematicity are also present. Indeed, these trends can
be understood theoretically as a result of the competition and coexistence
between magnetic/nematic order and superconductivity. On the one hand,
magnetism competes with superconductivity for the same electronic
states \cite{Chubukov2009,Fernandes2010}, which is most directly
revealed by the suppression of the magnetic order parameter below
$T_{c}$ seen in neutron scattering \cite{Avci2011}. Due to its magnetic
origin, nematic order inherits this competition and also competes
with superconductivity \cite{Moon10,Fernandes_Chubukov13}, as manifested
by the decrease of the orthorhombic distortion -- proportional to
the nematic order parameter \cite{Fernandes2014} -- below $T_{c}$,
as measured by x-ray scattering \cite{Avci2011}. The competition
between these electronic ordered states results in a suppression of the
zero-temperature superfluid density, and a consequent enhancement
of the penetration depth in the low-temperature limit \cite{Fernandes_Schmalian2,Vavilov_penetration_depth,Levchenko13,Sachdev13},
in qualitative agreement with the experimental data.

On the other hand, coexistence with magnetic/nematic order leads to
strong anisotropies in the gap function, which is also in qualitative
agreement with the experimental data. Consider for instance a simplified
scenario in which the $s^{+-}$ gap function and the pairing interaction
are completely isotropic at optimal doping. Because nematic order
breaks the tetragonal symmetry of the system, it gives rise to a d-wave
component in the original $s^{+-}$ gap function in the coexistence
state \cite{Fernandes_Millis13,Livanas12}. Due to the proximity between
the d-wave and $s^{+-}$ ground state energies -- as manifested by
the existence of a Bardasis-Schrieffer mode in the Raman spectrum
of the optimally doped samples \cite{raman_mode} -- this mixing between
d-wave and $s^{+-}$ states can be sizeable, leading to strong anisotropies
in the gap. Furthermore, long-range magnetic order promotes anisotropy
in the pairing interaction itself \cite{Chubukovanomaly}. Due to
the anisotropic reconstruction of the Fermi surface caused by the
doubling of the unit cell in the magnetic phase, the electronic states
near the Fermi level acquire a significant angular dependence, which
is translated to an effectively anisotropic pairing interaction for
the states of the reconstructed Fermi surface. As a result, the gap
nodes that were fully isotropic in the non-coexistence state develop
deep minima, which may even give rise to nodal behavior \cite{Chubukovanomaly}.

\begin{figure}[tb]
\includegraphics[width=0.8\linewidth]{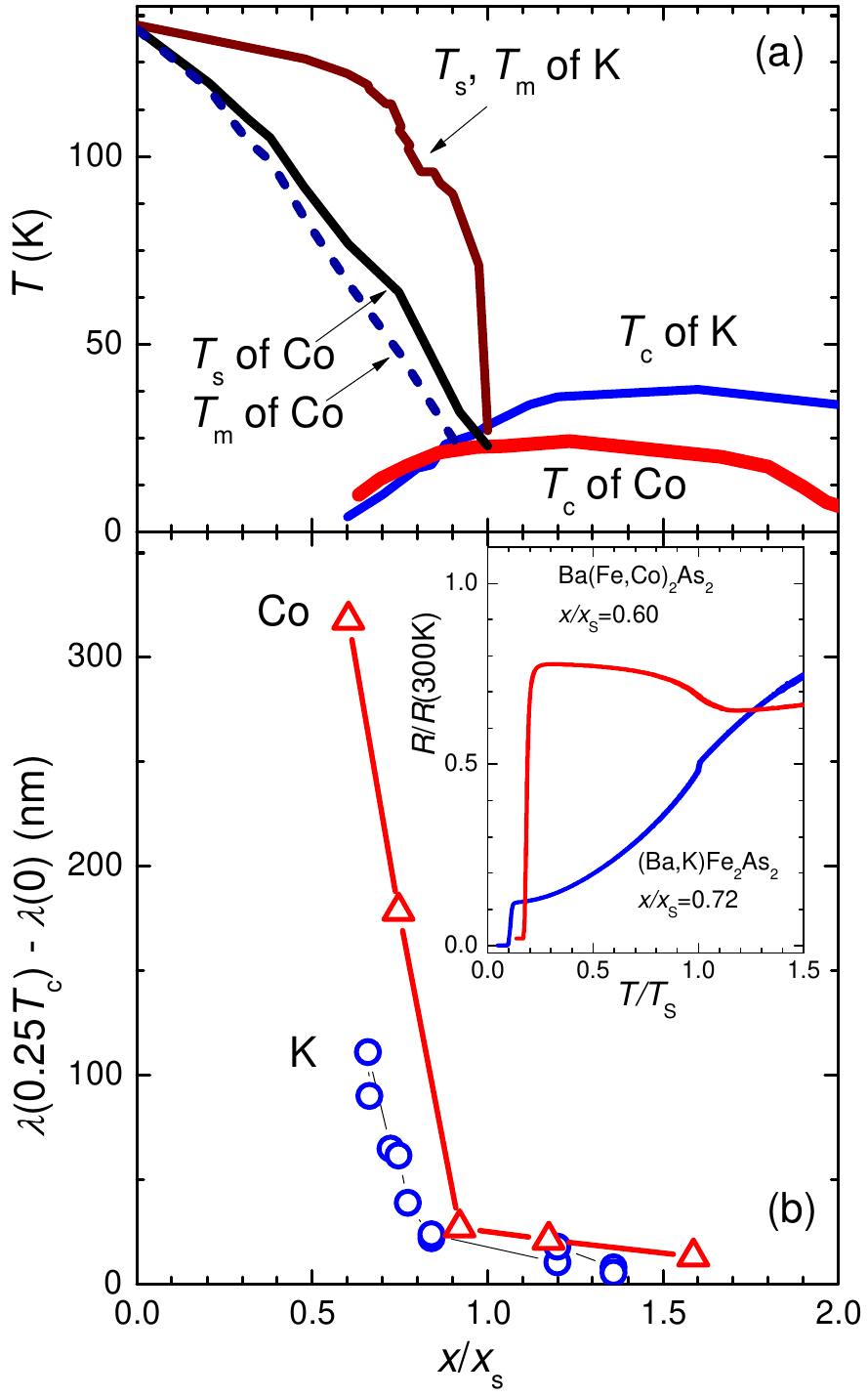} \caption{(a) Comparison of the doping phase diagrams of hole-doped Ba$_{1-x}$K$_{x}$Fe$_{2}$As$_{2}$
and of the electron-doped Ba(Fe$_{1-x}$Co$_{x}$)$_{2}$As$_{2}$.
The data are plotted using normalized $x/x_{s}$ composition scale,
where $x_{s}$ is a doping boundary of $T_{s}(x)$ lines with $x_{s}$=0.25
for K- and $x_{s}$=0.063 for Co-doping, respectively. (b) Doping
evolution of the temperature-dependent part of London penetration
depth $\Delta\lambda(0.25T_{c})$ in K-doped (blue circles) and Co-doped
(red triangles). Note three times difference in the magnitude of penetration
depth increase in two cases. Inset in panel (b) shows normalized temperature-dependent
resistivity, $R/R(\mathrm{300 K})$, for under-doped samples with Co-doping $x$=0.038, $x/x_{s}$=0.6
and K-doping $x$=0.18, $x/x_{s}$=0.72, plotted using normalized
temperature scale, $T/T_{s}$. Note that resistivity of K-doped samples
shows very small change at $T_{s}$, while that of Co-doped samples
increases significantly below $T_{s}$. }

\label{fig4}
\end{figure}

It is instructive to compare these observations with the results on
electron-doped BaCo122. In Fig.~\ref{fig4} we directly compare the
doping phase diagrams (panel (a)) and doping evolutions of the London
penetration depth (panel (b)) of the hole and electron-doped BaFe$_{2}$As$_{2}$
based superconductors. The data for BaCo122 were taken from Ref.~\onlinecite{Gordon2009}.
To take into account the fact that the structural (i.e. nematic) and
magnetic transition lines ($T_{s}$ and $T_{m}$, respectively) coincide
in BaK122, but split with doping in BaCo122, we normalized the compositions
of the samples to those with the lowest measured structural transition
temperatures, $x$=0.25 in BaK122 and $x$=0.063 in BaCo122. This
comparison reveals very interesting similarities and differences between
these two families. First, as can be seen from Fig.~\ref{fig4} the
$T_{s}(x)$ boundary in BaK122 terminates very sharply, suggesting
a possible first order transition with doping between the orthorhombic/nematic
and tetragonal phases. Second, the rapid increase of the London penetration
depth $\Delta\lambda(T_{up})$ with underdoping, reflecting the increase
of $\lambda(0)$ \cite{Gordon2010a}, has very different magnitudes
for the two types of doping. For instance, despite using BaK122 samples
that are much closer to the edge of the superconducting dome ($T_{c}$=7~K
for $x$=0.16) than in our previous study of BaCo122 compounds ($T_{c}$=7.4~K
for $x$=0.038) \cite{Gordon2010a}, we find a three times smaller
increase of $\Delta\lambda(0.25T_{c})$ on the hole-doped side compared
to the electron-doped side. This suggests weaker competition between
magnetic/nematic order and superconductivity in the hole-doped BaK122
than in the electron-doped BaCo122, which is in line with the significantly
weaker suppression of the magnetic order parameter below $T_{c}$
found in neutron scattering experiments \cite{Avci2011,Fernandes2010}.
Similarly, as shown in the inset of panel (b) in Fig.~\ref{fig4},
the normalized resistivity in the normal state is much less affected
by the long-range magnetic order in BaK122 than in BaCo122. This suggests
that the partial gapping of the Fermi surface by long-range magnetic
order is stronger in the latter case, leaving less electronic states
available for the superconducting state. We note that the nematic
susceptibility is also weaker in the BaK122 family, as evidenced by
the in-plane resistivity anisotropy behavior \cite{Blomberg2013}
and the softening of the shear modulus \cite{Boehmer2013}. Therefore,
our findings suggest a close relationship between the electron-hole
asymmetries of the normal state and superconducting properties. Despite
displaying different magnitudes, however, the continuous increase
of $\Delta\lambda(0.25T_{c})$ with underdoping in both electron-
and hole-doped samples contrasts with the case of isovalent doping,
BaP122, in which a sharp peak in the low-temperature penetration depth
is observed \cite{Hashimoto12}.

\section{Conclusions}

In conclusion, our measurements of the London penetration depth in
high quality single crystals of Ba$_{1-x}$K$_{x}$Fe$_{2}$As$_{2}$
close to the optimal doping level $x$=0.34 ($T_{c}=39$ K) reveal
a superconducting state with two full gaps $\Delta_{1}(0)$=6.5~meV
and $\Delta_{2}(0)$=3.3 meV. On the other hand, our measurements
deep in the underdoped regime, for $x$=0.16 ($T_{c}=7$ K), demonstrate
that the gap develops significant anisotropies without, however, developing
nodes. Comparison with the electron-doped compositions Ba(Fe$_{1-x}$Co$_{x}$)$_{2}$As$_{2}$
reveals a strong asymmetry of the structure of the superconducting
state, which is nodeless in hole-doped and nodal in electron-doped
compounds. These observations suggest that the competition and the
coexistence with magnetic/nematic order is responsible for the anisotropic
structure of the superconducting gap in the underdoped regime, and
that this competition is stronger in electron-doped rather than in
hole-doped compounds.

\section{ACKNOWLEDGMENTS}

We thank A. Chubukov, P. Hirschfeld, S. Maiti, J. Schmalian and L. Taillefer for
useful discussions. The work at Ames was supported by the U.S. Department of Energy (DOE),
Office of Science, Basic Energy Sciences, Materials Science and Engineering
Division. The research was performed at the Ames Laboratory, which
is operated for the U.S. DOE by Iowa State University under contract
DE-AC02-07CH11358. Work in China was supported by the Ministry of Science and Technology of China, project 2011CBA00102.
Work at Sherbrooke was supported by the Canadian Institute for Advanced Research and a Canada Research Chair, and it was funded by NSERC, FQRNT, and CFI.

\bibstyle{apsrev4-1}


\begin{thebibliography}{99}
\bibitem{Mazin2010Nature} I.~I. Mazin, Nature \textbf{464}, 183 (2010).

\bibitem{Wang2011} F. Wang and D.-H. Lee, Science \textbf{332}, 200
(2011).

\bibitem{reviews_pairing} P. J. Hirschfeld, M. M. Korshunov, and
I. I. Mazin, Rep. Prog. Phys. \textbf{74}, 124508 (2011); A. V. Chubukov,
Annu. Rev. Cond. Mat. Phys. \textbf{3}, 57 (2012).

\bibitem{Gordon2009} R. T. Gordon C. Martin, H. Kim, N. Ni, M. A.
Tanatar, J.~Schmalian, I. I. Mazin, S. L. Bud'ko, P. C. Canfield,
and R.~Prozorov, Phys. Rev. B \textbf{79}, 100506 (2009).

\bibitem{Gordon2009a} R. T. Gordon, N. Ni, C. Martin, M. A. Tanatar,
M. D. Vannette, H. Kim, G. D. Samolyuk, J. Schmalian, S. Nandi, A.
Kreyssig, A. I. Goldman, J. Q. Yan, S. L. Bud'ko, P. C. Canfield,
and R. Prozorov,, Phys. Rev. Lett. \textbf{102}, 127004 (2009).

\bibitem{Martin2010} C. Martin, H. Kim, R. T. Gordon, N. Ni, V. G.
Kogan, S. L. Bud'ko, P. C. Canfield, M. A. Tanatar, and R. Prozorov,,
Phys. Rev. B \textbf{81}, 060505 (2010).

\bibitem{Tanatar2010} M. A. Tanatar, J.-Ph. Reid, H. Shakeripour,
X. G. Luo, N. Doiron-Leyraud, N. Ni, S. L. Bud'ko, P. C. Canfield,
R. Prozorov, and Louis Taillefer, Phys. Rev. Lett. \textbf{104}, 067002
(2010).

\bibitem{Reid2010} J.-Ph. Reid, M. A. Tanatar, X. G. Luo, H. Shakeripour,
N. Doiron-Leyraud, N. Ni, S. L. Bud'ko, P. C. Canfield,R. Prozorov,
and Louis Taillefer, Phys. Rev. B \textbf{82}, 064501 (2010).

\bibitem{Gofryk2011} K. Gofryk A. B. Vorontsov, I. Vekhter, A. S.
Sefat, T. Imai, E. D. Bauer, J. D. Thompson, and F. Ronning, Phys.
Rev. B \textbf{83}, 064513 (2011).

\bibitem{Bud'ko2009} S. L. Bud'ko, Ni Ni, and Paul C. Canfield, Phys.
Rev. B \textbf{79}, 220516 (2009).

\bibitem{Hardy2010} F. Hardy, T. Wolf, R. A. Fisher, R. Eder, P.
Schweiss, P. Adelmann, H. v. L\"{o}hneysen, and C. Meingast,
Phys. Rev. B \textbf{81}, 060501 (2010).

\bibitem{Hirschfeld2010physics} P.~J. Hirschfeld and D.~J. Scalapino,
Physics \textbf{3}, 64 (2010).

\bibitem{Maiti2011} S. Maiti, M.M. Korshunov, T.A. Maier, P.J. Hirschfeld,
A.V. Chubukov, Phys. Rev. Lett \textbf{107}, 147002 (2011).

\bibitem{KyuilNaFeAs} K. Cho, M. A. Tanatar, N. Spyrison, H. Kim,
Y. Song, Pengcheng Dai, C. L. Zhang, and R. Prozorov, Phys. Rev. B
\textbf{86}, 020508 (2012).

\bibitem{M2s} R. Prozorov, K. Cho, H. Kim, and M.A.Tanatar, J. of
Physics Conference Series, \textbf{449}, 012020, (2013)

\bibitem{Borisenko2010} S.~V. Borisenko, V. B. Zabolotnyy, D. V.
Evtushinsky, T. K. Kim, I. V. Morozov, A. N. Yaresko, A. A. Kordyuk,
G. Behr, A. Vasiliev, R. Follath, and B. Buchner, Phys. Rev. Lett.
\textbf{105}, 067002 (2010).

\bibitem{Inosov2010} D.~S. Inosov, J. S. White, D. V. Evtushinsky,
I. V. Morozov, A. Cameron, U. Stockert, V. B. Zabolotnyy, T. K. Kim,
A. A. Kordyuk, S. V. Borisenko, E. M. Forgan, R. Klingeler, J. T.
Park, S. Wurmehl, A. N. Vasiliev, G. Behr, C. D. Dewhurst, and V.
Hinkov, Phys. Rev. Lett. \textbf{104}, 187001 (2010).

\bibitem{Kim2011} H. Kim, M. A. Tanatar, Yoo Jang Song, Yong Seung
Kwon, and R. Prozorov, Phys. Rev. B \textbf{83}, 100502 (2011).

\bibitem{Tanatar2011} M. A. Tanatar, J.-Ph. Reid, S. Ren\'{e} de Cotret, N. Doiron-Leyraud, F. Laliberte, E. Hassinger, J. Chang,
H. Kim, K. Cho, Yoo Jang Song, Yong Seung Kwon, R. Prozorov, and Louis
Taillefer, Phys. Rev. B \textbf{84}, 054507 (2011).

\bibitem{Chubukov2009} A.~V. Chubukov, M.~G. Vavilov, and A.~B.
Vorontsov, Phys. Rev. B \textbf{80}, 140515 (2009).

\bibitem{Hashimoto2010a} K. Hashimoto, M. Yamashita, S. Kasahara,
Y. Senshu, N. Nakata, S. Tonegawa, K. Ikada, A. Serafin, A. Carrington,
T. Terashima, H. Ikeda, T. Shibauchi, and Y. Matsuda, Phys. Rev. B
\textbf{81}, 220501 (2010).

\bibitem{Ding2008} H. Ding, P. Richard, K. Nakayama, T. Sugawara,
T. Arakane, Y. Sekiba, A. Takayama, S. Souma, T. Sato, T. Takahashi,
Z. Wang, X. Dai, Z. Fang, G. F. Chen, J. L. Luo, and N. L. Wang, ,
EPL \textbf{83}, 47001 (2008).

\bibitem{Nakayama2011} K. Nakayama, T. Sato, P. Richard, Y.-M. Xu,
T. Kawahara, K. Umezawa, T. Qian, M. Neupane, G. F. Chen, H. Ding,
and T. Takahashi, Phys. Rev. B \textbf{83}, 020501 (2011).

\bibitem{Li2011} Z. Li, D. L. Sun, C. T. Lin, Y. H. Su, J. P. Hu,
and Guo-qing Zheng, Phys. Rev. B \textbf{83}, 140506 (2011).

\bibitem{Martin2009} C. Martin, R. T. Gordon, M. A. Tanatar, H. Kim,
N. Ni, S. L. Bud'ko, P. C. Canfield, H. Luo, H. H. Wen, Z. Wang, A.
B. Vorontsov, V. G. Kogan, and R. Prozorov, Phys. Rev. B \textbf{80},
020501 (2009).

\bibitem{Hashimoto2009} K. Hashimoto, T. Shibauchi, S. Kasahara,
K. Ikada, S. Tonegawa, T. Kato, R. Okazaki, C. J. van der Beek, M.
Konczykowski, H. Takeya, K. Hirata, T. Terashima, and Y. Matsuda,
Phys. Rev. Lett. \textbf{102}, 207001 (2009).

\bibitem{Luo2009} X. G. Luo, M. A. Tanatar, J.-Ph. Reid, H. Shakeripour,
N. Doiron-Leyraud, N. Ni, S. L. Bud'ko, P. C. Canfield, Huiqian Luo,
Zhaosheng Wang, Hai-Hu Wen, R. Prozorov, and Louis Taillefer, Phys.
Rev. B \textbf{80}, 140503 (2009).

\bibitem{Dong2010} J. K. Dong, S. Y. Zhou, T. Y. Guan, H. Zhang,
Y. F. Dai, X. Qiu, X. F. Wang, Y. He, X. H. Chen, and S. Y. Li, Phys.
Rev. Lett. \textbf{104}, 087005 (2010).

\bibitem{Fukazawa2009} H. Fukazawa, T. Yamazaki, K. Kondo, Y. Kohori,
N. Takeshita, P. M. Shirage, K Kihou, K. Miyazawa, H. Kito, H. Eisaki,
and A. Iyo, J. Phys. Soc. Jpn. \textbf{78}, 033704 (2009).

\bibitem{Hashimoto2010} K. Hashimoto, A. Serafin, S. Tonegawa, R.
Katsumata, R. Okazaki, T. Saito, H. Fukazawa, Y. Kohori, K. Kihou,
C. H. Lee, A. Iyo, H. Eisaki, H. Ikeda, Y. Matsuda, A. Carrington,
and T. Shibauchi, Phys. Rev. B \textbf{82}, 014526 (2010).

\bibitem{Hardy2013} F. Hardy, A. E. Bohmer, D. Aoki, P. Burger, T.
Wolf, P. Schweiss, R. Heid, P. Adelmann, Y. X. Yao, G. Kotliar, J.
Schmalian, and C. Meingast, Phys. Rev. Lett. \textbf{111}, 027002
(2013).

\bibitem{Matsudaoverdoped} D. Watanabe, T. Yamashita, Y. Kawamoto,
S. Kurata, Y. Mizukami, T. Ohta, S. Kasahara, M. Yamashita, T. Saito,
H. Fukazawa, Y. Kohori, S. Ishida, K. Kihou, C. H. Lee, A. Iyo, H.
Eisaki, A. B. Vorontsov, T. Shibauchi, and Y. Matsuda, Phys. Rev.
B \textbf{89}, 115112 (2014).

\bibitem{ReidSUST} J.-Ph. Reid, A. Juneau-Fecteau, R. T. Gordon,
S. Rene de Cotret, N. Doiron-Leyraud, X. G. Luo, H. Shakeripour, J.
Chang, M. A. Tanatar, H. Kim, R. Prozorov, T. Saito, H. Fukazawa,
Y. Kohori, K. Kihou, C. H. Lee, A. Iyo, H. Eisaki, B. Shen, H.-H.
Wen, and Louis Taillefer, Supercond. Sci. Technol. \textbf{25}, 084013
(2012).

\bibitem{Fernandes2010} R. M. Fernandes, D. K. Pratt, Wei Tian,
J. Zarestky, A. Kreyssig, S. Nandi, Min Gyu Kim, A. Thaler, Ni Ni, P. C. Canfield, R. J. McQueeney, J. Schmalian, A. I. Goldman, Phys. Rev. B \textbf{81}, 140501 (2010).

\bibitem{Chubukovanomaly} S. Maiti, R. M. Fernandes, and A. V. Chubukov,
Phys. Rev. B \textbf{85}, 144527 (2012).

\bibitem{Fernandes2014} R. M. Fernandes, A. V. Chubukov, and J. Schmalian,
Nature Physics \textbf{10}, 97 (2014).

\bibitem{Blomberg2013} E. C. Blomberg, M. A. Tanatar, R. M. Fernandes,
I. I. Mazin, B. Shen, H.-H. Wen, M. D. Johannes, J. Schmalian, and
R. Prozorov, Nat. Commun. \textbf{4}, 1914 (2013).

\bibitem{Elena} E. Hassinger, G. Gredat, F. Valade, S. Rene de Cotret,
A. Juneau-Fecteau, J.-Ph. Reid, H. Kim, M. A. Tanatar, R. Prozorov,
B. Shen, H.-H. Wen, N. Doiron-Leyraud, and Louis Taillefer, Phys.
Rev. B \textbf{86}, 140502 (2012).

\bibitem{BaNa} S. Avci, J. M. Allred, O. Chmaissem, D. Y. Chung,
S. Rosenkranz, J. A. Schlueter, H. Claus, A. Daoud-Aladine, D. D.
Khalyavin, P. Manuel, A. Llobet, M. R. Suchomel, M. G. Kanatzidis,
and R. Osborn, Phys. Rev. B \textbf{88}, 094510 (2013).

\bibitem{Kogan2009} V. G. Kogan, C. Martin, and R. Prozorov, Phys.
Rev. B \textbf{80}, 014507 (2009).

\bibitem{Fernandes_Schmalian2} R. M. Fernandes and J. Schmalian,
Phys. Rev. B \textbf{82}, 014520 (2010).

\bibitem{Vavilov_penetration_depth} D. Kuzmanovski and M. G. Vavilov,
Supercond. Sci. Technol. 25, 084001 (2012).

\bibitem{Levchenko13} A. Levchenko, M. G. Vavilov, M. Khodas, and
A. V. Chubukov, Phys. Rev. Lett. \textbf{110}, 177003 (2013).

\bibitem{Sachdev13} D. Chowdhury, B. Swingle, E. Berg, and S. Sachdev,
Phys. Rev. Lett. \textbf{111}, 157004 (2013).

\bibitem{Boehmer2013} A. E. Bohmer, P. Burger, F. Hardy, T. Wolf,
P. Schweiss, R. Fromknecht, M. Reinecker, W. Schranz, and C. Meingast,
Phys. Rev. Lett. \textbf{112}, 047001 (2014).

\bibitem{Tanatar2010b}  M. A. Tanatar, N. Ni, A. Thaler, S. L. Bud'ko, P. C. Canfield and R. Prozorov, Phys. Rev. B  \textbf{82},  134528  (2010).

\bibitem{Tanatar2011BaKAniso}  M. A. Tanatar, W. E. Straszheim, H. Kim, J. Murphy, N. Spyrison, E. C. Blomberg, K. Cho, J.-P. Reid, B. Shen, L. Taillefer, H.-H. Wen and R. Prozorov, Phys. Rev. B  \textbf{89},  144514  (2014).

\bibitem{Luo2008} H. Q. Luo, Z. S. Wang, H. Yang, P. Cheng, X. Zhu,
and H.-H. Wen, Supercond. Sci. Technol. \textbf{21}, 125014 (2008).

\bibitem{Degrift1975} C. T. van Degrift, Rev. Sci. Instrum. \textbf{46},
599 (1975).

\bibitem{Prozorov2006} R. Prozorov and R. W. Giannetta, Supercond.
Sci. Technol. \textbf{19}, R41 (2006).

\bibitem{Prozorov2000} R. Prozorov, R. W. Giannetta, A. Carrington,
and F. M. Araujo-Moreira, Phys. Rev. B \textbf{62}, 115 (2000).

\bibitem{Prozorov2011} R. Prozorov and V. G. Kogan, Rep. Prog. Phys.
74, 124505 (2011).

\bibitem{Li2008} G. Li, W. Z. Hu, J. Dong, Z. Li, P. Zheng, G. F.
Chen, J. L. Luo, and N. L. Wang, Phys. Rev. Lett. \textbf{101}, 107004
(2008).

\bibitem{Mu2009Cp} G. Mu, Huiqian Luo, Zhaosheng Wang, Lei Shan,
Cong Ren, and Hai-Hu Wen, Phys. Rev. B \textbf{79}, 174501 (2009).

\bibitem{Rotter2008} M. Rotter, M. Tegel, and D. Johrendt, Phys.
Rev. Lett.\textbf{101}, 107006 (2008).

\bibitem{Avci2011} S. Avci, O. Chmaissem, E. A. Goremychkin, S. Rosenkranz,
J.-P. Castellan, D. Y. Chung, I. S. Todorov, J. A. Schlueter, H. Claus,
M. G. Kanatzidis, A. Daoud-Aladine, D. Khalyavin, and R. Osborn, Phys.
Rev. B \textbf{83}, 172503 (2011).

\bibitem{Moon10} E. G. Moon and S. Sachdev, Phys. Rev. B. \textbf{85},
184511 (2012).

\bibitem{Fernandes_Chubukov13} R. M. Fernandes, S. Maiti, P. Woelfle,
and A. V. Chubukov, Phys. Rev. Lett. \textbf{111}, 057001 (2013).

\bibitem{Fernandes_Millis13} R. M. Fernandes and A. J. Millis, Phys.
Rev. Lett. \textbf{111}, 127001 (2013).

\bibitem{Livanas12} G. Livanas, A. Aperis, P. Kotetes, and G. Varelogiannis,
arXiv:1208.2881.

\bibitem{raman_mode} F. Kretzschmar, B. Muschler, T. Bohm, A. Baum,
R. Hackl, H.-H. Wen, V. Tsurkan, J. Deisenhofer, and A. Loidl, Phys.
Rev. Lett. \textbf{110}, 187002 (2013).

\bibitem{Gordon2010a} R. T. Gordon, H. Kim, N. Salovich, R. W. Giannetta,
R. M. Fernandes, V. G. Kogan, T. Prozorov, S. L. Bud'ko, P. C. Canfield,
M. A. Tanatar, and R. Prozorov, Phys. Rev. B \textbf{82}, 054507 (2010).

\bibitem{Hashimoto12} K. Hashimoto, K. Cho, T. Shibauchi, S. Kasahara,
Y. Mizukami, R. Katsumata, Y. Tsuruhara, T. Terashima, H. Ikeda, M.
A. Tanatar, H. Kitano, N. Salovich, R. W. Giannetta, P. Walmsley,
A. Carrington, R. Prozorov, and Y. Matsuda, Science \textbf{336},
1554 (2012).\end{thebibliography}
\end{document}